\documentclass[iop, appendixfloats, numberedappendix]{emulateapj}

\usepackage[fleqn]{amsmath}

\usepackage{color}

\newcommand{\eps}{\ensuremath{\varepsilon}}

\newcommand{\kep}{Kepler-41}
\newcommand{\kepb}{Kepler-41b}
\newcommand{\kepf}{Kepler-41}
\newcommand{\kepfb}{Kepler-41b}

\newcommand{\kfb}{Kepler-41b}

\newcommand{\keps}{Kepler-7}
\newcommand{\kepsb}{Kepler-7b}
\newcommand{\ksb}{Kepler-7b}
 
\newcommand{\kept}{Kepler-12}
\newcommand{\keptb}{Kepler-12b}

\newcommand{\ktb}{Kepler-12b}
 
\newcommand{\ik}{{\it Kepler}}

\newcommand{\is}{{\it Spitzer}}

\newcommand{\sig}[1]{\ensuremath{#1\sigma}}

\newcommand{\figr}[1]{Figure~\ref{fig:#1}}

\newcommand{\secr}[1]{Section~\ref{sec:#1}}
\newcommand{\appr}[1]{Appendix~\ref{app:#1}}

\newcommand{\tabr}[1]{\mbox{Table~\ref{tab:#1}}}

\shorttitle{Evidence that inhomogeneous atmospheric reflection is common}

\shortauthors{Shporer \& Hu}

\begin{document}

\title{Studying atmosphere-dominated hot Jupiter Kepler phase curves:\\
Evidence that inhomogeneous atmospheric reflection is common.}

\author{Avi Shporer\altaffilmark{1, 2, 3}, 
Renyu Hu\altaffilmark{1, 2, 4},
} 
\altaffiltext{1}{Jet Propulsion Laboratory, California Institute of Technology, 4800 Oak Grove Drive, Pasadena, CA 91109, USA}
\altaffiltext{2}{Division of Geological and Planetary Sciences, California Institute of Technology, Pasadena, CA 91125, USA}
\altaffiltext{3}{NASA Sagan Fellow}
\altaffiltext{4}{NASA Hubble Fellow}

\begin{abstract}

We identify three \ik\ transiting planets, \kepsb, \keptb, and \kepfb, whose orbital phase-folded light curves are dominated by planetary atmospheric processes including thermal emission and reflected light, while the impact of non-atmospheric (i.e.~gravitational) processes, including beaming (Doppler boosting) and tidal ellipsoidal distortion, is negligible. Therefore, those systems allow a direct view of their atmospheres without being hampered by the approximations used in the inclusion of both atmospheric and non-atmospheric processes when modeling the phase curve shape. We present here the analysis of \keptb\ and \kepb\ atmosphere based on their \ik\ phase curve, while the analysis of \kepsb\ was already presented elsewhere. The model we used efficiently computes reflection and thermal emission contributions to the phase curve, including inhomogeneous atmospheric reflection due to longitudinally varying cloud coverage. We confirm \keptb\ and \kepb\ show a westward phase shift between the brightest region on the planetary surface and the substellar point, similar to \kepsb. We find that reflective clouds located on the west side of the substellar point can explain the phase shift. The existence of inhomogeneous atmospheric reflection in all three of our targets, selected due to their atmosphere-dominated \ik\ phase curve, suggests this phenomenon is common. Therefore it is likely to be present also in planetary phase curves that do not allow a direct view of the planetary atmosphere as they contain additional orbital processes. We discuss the implications of a bright-spot shift on the analysis of phase curves where both atmospheric and gravitational processes appear, including the mass discrepancy seen in some cases between the companion's mass derived from the beaming and ellipsoidal photometric amplitudes. Finally, we discuss the potential detection of non-transiting but otherwise similar planets, whose mass is too small to show a gravitational photometric signal but their atmosphere is reflective enough to show detectable phase modulations.

\end{abstract}

\keywords{planetary systems --- stars: individual (Kepler-7, Kepler-12, Kepler-41) --- techniques: photometric}


\section{Introduction}
\label{sec:intro}

The availability of space-based high-quality time series photometry over the last several years has allowed the monitoring of transiting planetary systems not only during transit or occultation (secondary eclipse) but throughout their entire orbit, for a growing sample of planets. This was done at first by \is\ in the infrared \citep[e.g.,][]{knutson07, knutson09, knutson12, cowan12, lewis13}, thus facilitating an improved understanding of processes taking place in planetary atmospheres, such as winds that shift the hottest region on the planet surface eastward from the substellar point \citep{showman02}. Later, the {\it CoRoT} and \ik\ space missions enabled similar monitoring in visible light (optical), which is sensitive also to light from the host star reflected by the planetary atmosphere. 

Moreover, such high-quality space-based optical data is sensitive also to non-atmospheric processes induced by the planet-star gravitational interaction \citep[e.g.,][]{mazeh10, shporer11, faigler11}, including beaming \citep[aka Doppler boosting; e.g.,][]{loeb03, zucker07, shporer10, vankerkwijk10, bloemen11} and tidal ellipsoidal distortion modulations \citep[which we refer to hereafter simply as ellipsoidal; e.g.,][]{morris85, morris93}. Since both processes are gravitational in origin they are both sensitive to the orbiting companion's mass. Therefore, phase curves in the optical allow the study of both the planetary atmosphere and the gravitational interaction between the planet and the host star.

\citet{faigler11} proposed to utilize phase curve modulations of non-transiting systems to infer the existence of an unseen orbiting companion (substellar or stellar), by simultaneously modeling all processes, atmospheric and gravitational, and using the latter to estimate the companion's mass. Their simplistic model consists of simple approximations for each process, and their approach has already proven successful (\citealt{faigler12, faigler13}; see also \citealt{shporer11}).

Recently, the number of transiting planets with a measured optical phase curve has grown significantly \citep[e.g.,][]{esteves13, angerhausen14}, and the existence of a shift between the substellar point and the brightest region on the planet surface was suggested for a few of them \citep{demory13, esteves15, faigler15}. Such shifts include both eastward and westward bright-spot shifts, corresponding to a pre-occultation and a post-occultation maximum (respectively) in the phase curve's atmospheric component.

However, analysis of the increasing sample of phase curves has resulted in a few inconsistencies. Most notably is a discrepancy that appears in some cases between the companion's mass derived from the beaming photometric amplitude and the mass derived from the ellipsoidal photometric amplitude. For systems where radial velocities (RVs) of the host star are available the RV mass confirms the beaming mass in some cases while in others it confirms the ellipsoidal mass. This {\it mass discrepancy} appears in both star-planet systems, e.g., Kepler-13A \citep{shporer11, mazeh12, esteves13, shporer14}, TrES-2 \citep{barclay12}, and HAT-P-7 \citep{esteves13}, and stellar binaries with a transiting white dwarf companion, including KOI-74 \citep{vankerkwijk10, bloemen12}, and KIC 10657664 \citep{carter11}. This points to a gap in the understanding of one or more of the processes governing optical phase curves. 

Confronting the mass discrepancy described above can be done by studying phase curves showing only one of the processes governing their shape. For the beaming effect this was done by \cite{shporer10} who measured the optical (SDSS-$g$ band) phase curve of a detached eclipsing double white dwarf binary \citep{steinfadt10} where the objects' compact nature results in only the beaming effect being detectable at a high signal to noise ratio. Here we carry out a study of planetary atmospheres using \ik\ phase curves that are completely dominated by atmospheric processes, including both thermal emission and reflected light, while the gravitational processes (beaming and ellipsoidal) are much smaller in photometric amplitude, by an order of magnitude or more, and are close to the noise level. Therefore, such phase curves present an opportunity for studying planetary atmospheres as they are clean in the sense that they allow a careful study of the planetary atmosphere without being hampered by the approximations done in the simultaneous modeling of all processes affecting the phase curve shape \citep{faigler11}. For the atmosphere-dominated phase curves we identify we use the formalism of \cite{hu15} for a more detailed study of the planetary atmospheres than done previously.

Out of the $\approx$20 currently known hot Jupiters with a measured mass in the \ik\ field we have identified three systems whose phase curve is dominated by atmospheric processes: Kepler-7 \citep{latham10, demory11}, Kepler-12 \citep{fortney11}, and Kepler-41 \citep{santerne11, quintana13}. For those three systems the gravitational processes are at the 1 part per million (ppm) level in relative flux, which is more than an order of magnitude smaller than the observed phase curve amplitude and well within the phase curve noise level. Another five systems include significant non-atmospheric processes in their phase curve (TrES-2, HAT-P-7, Kepler-13, Kepler-76, Kepler-412). For the rest a phase curve signal was not detected, either due to insufficient signal to noise ratio, or due to stellar activity, from the planet host star or another star within the \ik\ pixel mask, that prevents identifying the orbital signal.

We describe our \ik\ data analysis and model fitting in Section~2 and show our results in Section~3. In Section~4 we discuss our findings and their implications, and bring a short summary in Section~5.

\section{Data Analysis}
\label{sec:dataanal}

\subsection{Data preprocessing}
\label{sec:prepro}

We use here \ik\ long cadence data from quarters~2 through 17\footnote{\kept\ was located on \ik\ CCD module~3 during quarter~4, when that module failed. So it is missing part of quarter~4 and the entire quarters~8, 12, and 16.}. Quarters~0 and 1 are ignored since they show a larger scatter than the other quarters. We have applied several preprocessing steps to the data, following \cite{shporer11, shporer14}, to prepare it for the atmospheric model fitting. We first remove instrumental signals, or trends, by fitting the first four cotrending basis vectors (CBVs) to the data of each quarter using the Pyke Python package \citep{still12}. We checked that increasing to six CBVs gave consistent results well within \sig{1}, consistent with \cite{shporer11, shporer14} and \cite{hu15}.

We then fit each continuous data segment with a 5th degree polynomial while ignoring in-transit and in-occultation data, and then dividing all data within the segment by that polynomial. This step does not affect the sinusoidal modulation along the orbit since the duration of each continuous segment is at least an order of magnitude longer than the orbital period. Fitting was done while iteratively rejecting \sig{5} outliers until none are identified using the formalism described in \citet[][see their Section 2]{shporer14}. Using polynomial of degrees 4 and 6 did not change the results. Using smaller degrees showed an increased overall scatter in the light curves while showing the same sinusoidal amplitude, where the latter is the astrophysical signal we seek to characterize here. Using larger degrees decreased the sinusoidal amplitude resulting from the excessively aggressive detrending. The phase curves before and after the polynomial detrending are presented in \appr{polycomp}.

Next we phase folded each light curve using the known ephemeris and binned it using 200 phase bins. We rejected \sig{4} outliers from each bin, which contained about 230 and 300 individual \ik\ measurements for \kept\ and \kepf, respectively. For each bin we calculated the median relative flux value, and the median absolute deviation (also known as MAD) times 1.4826 to be the uncertainty. Therefore the bins' uncertainties are independent of the error bars assigned to the \ik\ individual measurements and is based on the true scatter in each phase bin. We compared the measured scatter within each phase bin to the expected scatter based on the error bars of the individual measurements and Poisson statistics. We found that on average the measured scatter is 11\% and 49\% larger than the expected scatter for \kept\ and \kepf, respectively. For \kepf\ the increased measured scatter can be at least partially attributed to stellar activity (see Section~4.3). Although, for \kept\ there is no evidence for stellar activity (see Section~4.3) so the increased measured scatter is attributed to residual correlated noise, which in principle may exist also in the \kepf\ data but for the latter the stellar activity signal dominates. 

Next we checked whether combining the data from different \ik\ quarters follows Poisson statistics or whether it adds some noise due to, e.g., instrumental differences between quarters or data detrending. For each object we generated a phase folded and binned light curve for each quarter. We then calculated for each phase bin the ratio between the average uncertainty, averaging across all quarters, and the uncertainty of the phase bin when using all \ik\ data, and further dividing  by the square root of the number of quarters. For both the \kept\ and \kepf\ data sets the average ratio across all phase bins was within 1--2 \% from unity, showing that combining data from different quarters closely follows Poisson statistics.

Finally, we checked that our results do not depend on the phase bin width by analyzing also phased light curves with 100 and 400 bins and confirming the results are identical. 

Before moving to the atmospheric modeling we removed from the phase-folded light curves the orbital modulations induced by beaming and ellipsoidal, based on the known planet mass in each system \citep{santerne11, demory11}. This is done for completeness since as mentioned above, for the systems analyzed here these modulations are small, at the noise level, and the phase curve amplitude is dominated by atmospheric processes (see \figr{lcmodel}) so any approximations in determining the shape of the beaming and ellipsoidal phase curves do not affect the overall phase curve shape significantly.

\subsection{Phase curve analysis with atmospheric modeling}
\label{sec:atm}

We analyze the \ik\ phase curves using the interpretation framework described in \cite{hu15}. The model efficiently computes reflection and thermal emission contributions to the phase curve, considering both hot spot shift due to equatorial winds and inhomogeneous atmospheric reflection due to patchy clouds. The model assumes the atmosphere has clear longitudes and cloudy longitudes, and the cloudy part is more reflective than the clear part. The model assumes the cloud distribution is controlled by the temperature distribution, which is in turn approximated by the analytical model proposed by \cite{cowan11}. The model also assumes the thermal emission comes from a thermal photosphere that may be hotter than the equilibrium temperature. Here, the equilibrium temperature, as a function of longitude, is defined to be the atmospheric temperature without any greenhouse effects, and is calculated from the Bond albedo and the heat redistribution efficiency.

We fit the phase curve to this model characterized by five parameters: the Bond albedo, a heat redistribution efficiency, a greenhouse factor, the condensation temperature of clouds, and a reflectivity boosting factor by clouds. The Bond albedo describes the overall reflectivity of the planet, and takes any value between 0 and 1. The heat redistribution efficiency, the ratio between the radiative timescale and the advective timescale, controls the longitudinal temperature distribution. The magnitude of this parameter describes how well heat is transported: the larger the parameter, the smaller the longitudinal temperature variation becomes. The sign of this parameter describes the direction of the equatorial winds: a positive sign indicates super-rotating eastward winds, and a negative sign indicates westward winds. The greenhouse factor is the ratio between the temperature of the thermal photosphere and the equilibrium temperature, and takes any value not smaller than 1. The cloud condensation temperature, another free parameter, determines the longitudinal boundaries of the clouds. Finally, the reflectivity boosting factor by clouds is the proportional increase in reflectivity of an atmospheric patch when it becomes cloudy, and takes any positive value. The rest of the system parameters are known and taken from the literature \citep{fortney11, santerne11, quintana13}.

We use the Markov-Chain Monte Carlo (MCMC) method \citep{haario06} to explore the parameter space. Following \cite{gelman92} we calculate two Markov chains for each of the two objects, each chain containing $10^6$ steps where the first half of each chain is considered the ``burn-in'' period and is removed once the chain is completed. We verify that the R values for all parameters are less than 1.01 to ensure convergence \citep{gelman92}, and that in both cases the results from the two chains are identical. The allowed ranges and the prior distributions of the model parameters are as follows: the Bond albedo uniformly ranges in [0,1]; the heat redistribution efficiency uniformly ranges in [0,100]; the greenhouse factor uniformly ranges in [1,2]; the cloud condensation temperature uniformly ranges in [1000, 3000] K; and the reflection boosting factor uniformly ranges in [0,100]. The redistribution efficiency parameter is only allowed to take a positive value because atmospheric circulation models predict eastward equatorial winds \citep[e.g.,][]{showman02, showman13}. This choice effectively requires any post-occultation phase offset to be explained by the asymmetric reflection components induced by the patchy clouds, rather than the thermal emission component \citep{hu15}. Based on the converged Markov chain, we derive the occultation depth, the phase curve amplitude, and the phase shift of the phase curve maximum from occultation phase. The latter is defined to be positive for post-occultation maximum. 

After carrying out the analysis once, we noticed that for \keptb\ the reduced $\chi^2$ of the residuals was 1.12. Therefore we repeated the analysis of that phase curve while increasing the bins' uncertainties to bring the reduced $\chi^2$ to unity. This has increased the uncertainty on the fitted parameters by up to a few percent. For \kepfb\ the reduced $\chi^2$ was 0.86 in the first analysis, so no further analysis was done.

To test our results we applied the so-called ``prayer bead" approach to both objects, where the residuals are cyclicly permuted and added back to the model, and the new data set is then refitted. This approach preserves any correlated noise features in the data and produces a distribution for each fitted parameter from which its uncertainty can be estimated \citep[e.g.,][]{bouchy05, southworth08, winn08}. For both objects the prayer bead analysis did not result in larger uncertainties than the original MCMC analysis, so we use the latter as our final result. 

\section{Results}
\label{sec:res}

Our resulting fitted light curve models are shown in \figr{lcmodel} and the fitted and derived parameters are listed in \tabr{fitparams}. In the latter we list also the results for \kepsb\ from \citet{hu15} for comparison and completeness. In \appr{matfig} we show the correlations between the five fitted model parameters for each of the two objects.

For both \keptb\ and \kepfb\ we find that their phase curves are dominated by atmospheric reflection and we identify a statistically significant shift of the phase curve maximum from the occultation phase where the maximum is post-occultation, similar to \kepsb\ \citep{demory13, hu15, webber15}. 

\subsection{\keptb}
\label{sec:k12}

We measure a phase curve amplitude of $17.5\pm2.2$ ppm (part per million) and an eclipse depth of $14.3\pm2.5$ ppm, consistent with previous studies \citep{angerhausen14, esteves15} but with smaller uncertainties. We detect a clear phase curve asymmetry, with a phase curve maximum shifted by $47.6^{+6.9}_{-7.6}$ deg from occultation phase. This is consistent with the phase shift reported by \cite{esteves15} but $\approx\sig{2}$ smaller than that of  \citet[][D.~Angerhausen, private communication]{angerhausen14}. We find that the phase curve amplitude is greater than the eclipse depth by $\approx$$\sig{1}$, in agreement with the measurements of \cite{esteves15}. The difference between the phase curve amplitude and eclipse depth is enhanced by the relatively large phase offset.

To further test our analysis we carried out a separate analysis where we fitted a simple parametric model including a sinusoidal amplitude and phase. This is similar to previous analyses (see a summary of previous results in Table~9 of \citealt{esteves15}). We derived a peak to peak amplitude of $19.0 \pm 2.2$ ppm and a phase of maximum shifted later than the occultation phase by $58.5 \pm 6.8$ deg. These values are $\lesssim$ \sig{1} away from that of our original analysis and with similar uncertainties (see \tabr{fitparams}). 

\begin{figure*}
\begin{center}
\includegraphics[scale=0.505]{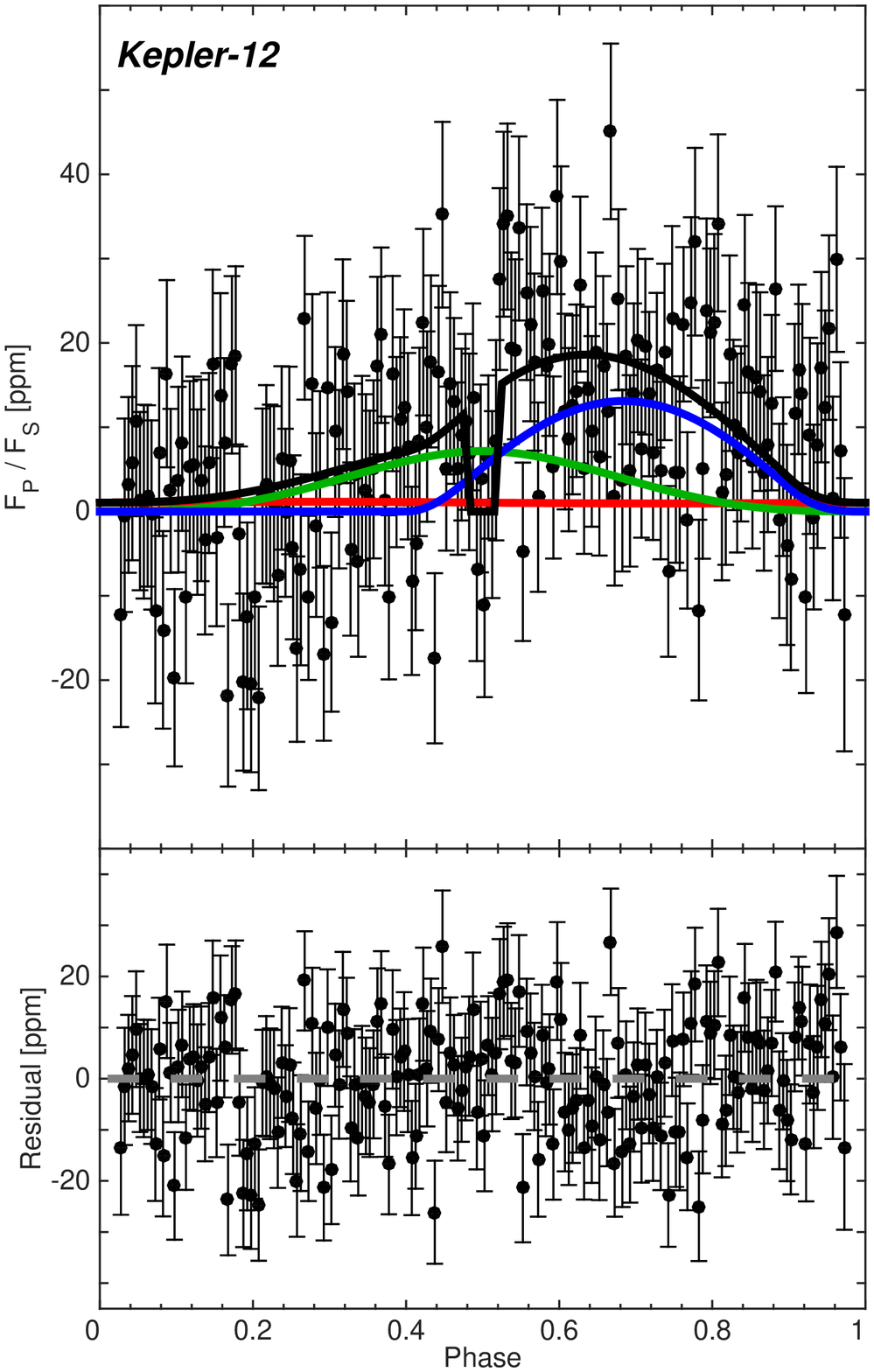}
\includegraphics[scale=0.505]{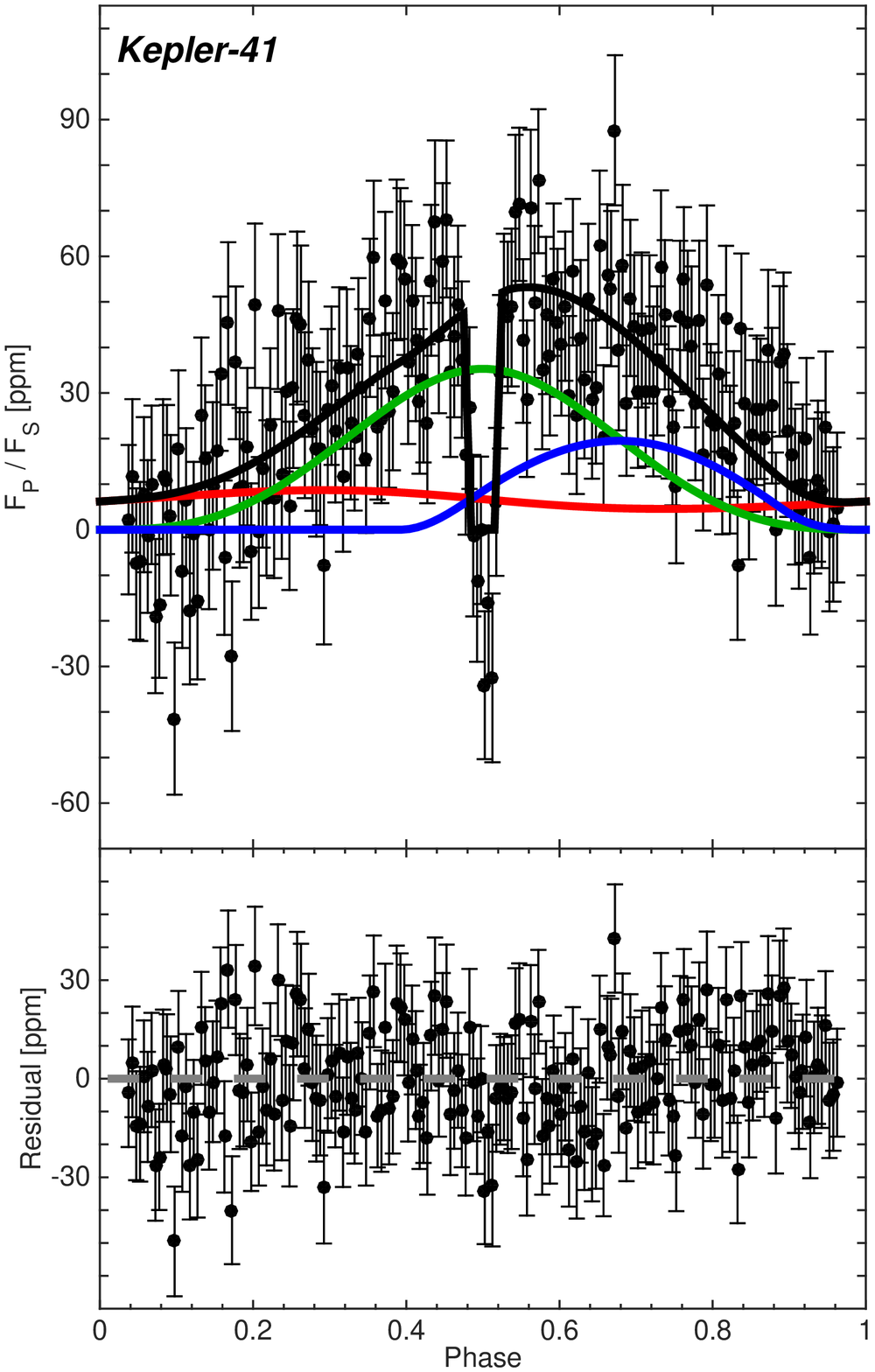}
\caption{\label{fig:lcmodel} Phase folded and binned \ik\ light curves of \keptb\ (left) and \kepfb\ (right). Top panels show relative flux (planet flux, $F_p$, relative to the in-occultation stellar flux, $F_s$) in part per million (ppm) versus orbital phase. Occultation (secondary eclipse) is at phase 0.5 and transit (not shown) at 0.0. The black curve shows the best-fit model phase curve and the colored lines show contribution of thermal emission (red), symmetric reflection (green), and asymmetric reflection (blue). The model phase curve has a significant contribution from the asymmetric reflection component due to patchy clouds. The clouds concentrate on the west side of the substellar point, due to a hot spot shifted eastward. The bottom panels show the residuals, data subtracted by the model, with a gray dashed line at zero residual for reference. For these systems the orbital gravitational processes, beaming and ellipsoidal, are typically at the 1 ppm level, that is close to the noise level and much smaller than the overall phase curve amplitude which is dominated by atmospheric processes. 
}
\end{center}
\end{figure*}

Looking at \keptb\ phase curve (\figr{lcmodel} left panel), there are a few consecutive bins around phase 0.2 that fall below the fitted model. While it could be an astrophysical signal that is not included in our model, it could also be a statistical, correlated noise feature. Although, correlated noise is expected to average out in the phase folded light curve. Either way, as described in \secr{atm} we have inflated the bins' uncertainties to bring the residuals reduced $\chi^2$ to unity, and the prayer bead analysis did not result in  larger error bars for the fitted parameters compared to the MCMC analysis.

\subsection{\kepfb}
\label{sec:k41}

Although an asymmetry in \kepfb\ phase curve is not identified by \cite{angerhausen14}, it is identified by \cite{esteves15} and is visually apparent in \citet[][their Fig.~3]{santerne11} and \citet[][their Fig.~2]{quintana13}. Here we confirm the post-occultation phase maximum which we measure to be shifted by $17.8 \pm6.3$ deg. We derive an eclipse depth of $49.1^{+6.4}_{-5.2}$ ppm and a phase amplitude of $48.7 \pm 5.2$ ppm, within \sig{$1--2$} from previous results but with typically smaller error bars \citep{santerne11, quintana13, angerhausen14, esteves15}. \citet{esteves15} report a phase amplitude greater than the eclipse depth (although at low statistical significance of only \sig{1.2}), while here we find a phase curve amplitude that is highly consistent (within \sig{0.1}) with the eclipse depth, indicating either poor heat redistribution or large planetary albedo.

We fitted \kepfb\ with the same simple parametric model as we did for \keptb, consisting of a sinusoidal amplitude and phase. We derived a peak to peak sinusoidal amplitude of $39.2 \pm 4.2$ ppm, and a phase of maximum shifted later than the occultation phase by $25.2 \pm 5.9$ deg. These are within \sig{1.5} from the results of our original analysis and with similar uncertainties (see \tabr{fitparams}).

\begin{deluxetable*}{llll}
\tablecaption{\label{tab:fitparams} } 
\tablewidth{0pt}
\tablehead{\multicolumn{1}{l}{Parameter} & \multicolumn{1}{l}{\keptb}  & \multicolumn{1}{l}{\kepfb} &  \multicolumn{1}{l}{\kepsb $^6$}  }
\startdata
Fitted parameters: & & \\
Bond albedo &$0.12\pm0.02$ &$0.25\pm0.06$ & $0.42\pm0.01$      \\
Heat redistribution & $46^{+34}_{-27}$ & $2.2^{+36.1}_{-1.3}$  & $49^{+33}_{-29}$ \\
Greenhouse factor & $1.05\pm0.05$ & $1.06^{+0.11}_{-0.04}$  & $1.08^{+0.08}_{-0.05}$       \\
Cloud condensation temp. [K]  & $1506^{+13}_{-24}$ & $1580^{+150}_{-120}$ & $1480\pm10$  \\
Cloud reflectivity factor & $45^{+36}_{-27}$ & $27^{+48}_{-23}$  & $28^{+41}_{-15}$     \\
\hline
Derived Parameters: & & \\
Eclipse depth [ppm] & $14.3\pm2.5$ & $49.1^{+6.4}_{-5.2}$ & $39.0^{+1.8}_{-1.6}$  \\
Phase amplitude [ppm] & $17.5\pm2.2$ & $48.7\pm5.2$ & $46.6\pm1.3$ \\
Phase offset [deg] $^1$ & $47.6^{+6.9}_{-7.6}$ & $17.8\pm6.3$ & $36.2\pm2.7$  \\
West longitude [deg] $^2$ & $-33\pm15$ & $-32.1^{+31.3}_{-12.3}$ & $-11.2\pm2.7$  \\
East longitude [deg] $^3$ & 90 & 90 & 90   \\
$A_{clear}$ $^4$  & $0.0097^{+0.0191}_{-0.0051}$ & $0.026^{+0.174}_{-0.017}$ & $0.034^{+0.038}_{-0.020}$  \\
$A_{cloud}$ $^5$ &  $0.44^{+0.27}_{-0.14}$ &  $0.68^{+0.21}_{-0.31}$ & $0.92\pm0.04$  
\enddata
\tablecomments{\\
1: Defined to be positive for post-occultation maximum. \\
2: Longitudinal boundary of clouds in the western hemisphere, with the substellar point defined as zero longitude. \\
3: Longitudinal boundary of clouds in the eastern hemisphere. 90 degree means the eastern hemisphere is completely cloud-free.\\
4: Reflectivity of the clear longitudes. \\
5: Reflectivity of the cloudy longitudes. \\
6: Results from \citet{hu15}.
}
\end{deluxetable*}

\section{Discussion}
\label{sec:dis}

\subsection{Atmospheric constraints}

Our analysis of \keptb\ and \kepb\ \ik\ phased light curves confirms and refines previous results and includes a more detailed atmospheric analysis than done previously.

We identify an asymmetry in the \ik\ phased light curves for the two planetary systems studied here, where the maximum is post-occultation. Our interpretation is that this asymmetry is due to the brightest region on the planetary surface being shifted westward from the substellar point, that in turn originates from an asymmetric surface distribution of clouds in the planetary atmosphere. The post-occultation maximum requires that the reflective clouds exist on the west side of the substellar point but not on the east side. Such asymmetric cloud distribution may be a result of eastward heat transport driven by zonal jets in the atmosphere. In addition, the clear part of the atmosphere must be quite dark, probably due to absorption by Alkali metal vapors deep in the atmosphere, whereas the cloudy part appears to be bright due to Mie scattering \citep{sudarsky00, sudarsky03}. 

For \kepfb, in order to produce a phase variation consistent with \ik\ observations, the east side of the sub-stellar meridian should be devoid of reflective clouds and have a reflectivity less than 0.2 (\sig{1} limit, see Table~\ref{tab:fitparams}). \figr{lcmodel} shows that the thermal emission component has the opposite phase offset than the asymmetric reflection component. On the west side, where reflective clouds are present, the reflectivity should be greater than 0.3. The overall Bond albedo of the planet is constrained to be $0.25 \pm 0.06$. The cloud condensation temperature parameter is found to be $1580^{+150}_{-120}$ K. To compare with material properties, one should also consider that clouds can form deep in the atmosphere and hence at a higher temperature \citep{hu14}. After multiplying by the derived greenhouse factor the \sig{1} cloud condensation temperature range is 1530--1910 K. Silicates (MgSiO$_3$ and Mg$_2$SiO$_4$) are consistent with this temperature range and are highly reflective, so are therefore candidate condensates.

Comparing Kepler-12b with Kepler-41b, we find that Kepler-12b is a more extreme case. Its phase curve is completely dominated by asymmetric surface distribution of reflective clouds, whereas the phase curve of Kepler-41b is dominated by symmetric reflection (see \figr{lcmodel}). Because the cloud distribution is controlled by the cloud condensation temperature, the post-occultation phase offset yields a tight constraint on the latter ($1506^{+13}_{-24}$ K). Multiplying by the derived greenhouse factor the \sig{1} cloud condensation temperature range is 1480--1670 K. To produce the phase offset, the clear part of \keptb\ atmosphere needs to have an albedo very close to zero, lower than that of Kepler-41b. This can be naturally explained as a higher metallicity and therefore greater Na and K absorption in the atmosphere of Kepler-12b \citep{sudarsky00, sudarsky03}. However, the host star in the Kepler-12 system is relatively metal-poor compared to Kepler-41 \citep{fortney11, santerne11}, therefore it is unclear why Kepler-12b would have a more metal-rich atmosphere than Kepler-41b. 

Overall, our analysis shows that it is reasonable to assume that an asymmetric cloud distribution is a common phenomena for hot Jupiters, but the details for the clear and cloudy parts of their atmospheres can still manifest great diversities.

For both planetary atmospheres the current data constrain the Bond albedo and the cloud condensation temperature, and put an upper limit on the greenhouse factor. Among these parameters, we see that the Bond albedo and the cloud condensation temperature are somewhat correlated, as expected (See Figures~\ref{fig:corr1} and \ref{fig:corr2}). This is because to produce a patchy cloud solution the cloud condensation temperature needs to be between the maximum and the minimum temperature of the dayside, both affected by the Bond albedo. The same correlation is observed for \kepsb\ \citep{hu15}. We note that the efficiency of heat redistribution cannot be sufficiently constrained by the current data.  Although the best-fit models suggest efficient heat redistribution, many more models would be permitted at an expense of minimally increasing $\chi^2$. For example, we find that much less efficient heat redistribution and a much greater reflectivity contrast between the cloudy part and the clear part could render an almost-as-good fit for \kepfb. As discussed in \cite{hu15}, this insensitivity to the exact value of the heat redistribution efficiency is tied to the uncertainty in the exact condensation temperature of the cloud material.

The atmospheric analysis done here for \keptb\ and \kepfb\ and the analysis done for \kepsb\ by others \citep{demory13, hu15, webber15} point to a similar scenario, that the optical \ik\ phase curve is dominated by reflected light (as opposed to thermal emission) and that the shift between the phase curve maximum and planetary occultation phase is a result of a non-uniform longitudinal cloud coverage. The irradiation received by these three planets is similar, and hence the cloud condensation temperatures derived from our phase curve analysis are also similar. To form the patchy clouds prominent longitudinal distribution the condensation/evaporation timescale of the cloud particles needs to be shorter than their advection timescale \citep{hu15}. This condition is uniquely met by hot Jupiters that have fast zonal winds in their atmospheres \citep[e.g.][]{knutson07}.

\subsection{Phase shift implications}

The three planets identified here as having a phase curve dominated by atmospheric processes, hence they allow a direct view of their atmospheres, all show a phase shift in the form of a post-occultation maximum. The identification of phase shifts in these atmosphere-dominated phase curves suggests they are common and exist also in phase curves of other hot Jupiters where they are more difficult to identify uniquely as those systems show additional phase modulations, induced by beaming and tidal ellipsoidal distortion. This has several potential implications we discuss below.

One obvious implication of the phase shift is on atmospheric studies using only the occultation and not the entire phase curve. Such studies commonly assume that the occultation depth is a flux measurement of the brightest region on the planetary surface, which is not the case for shifted phase curves. In those cases, the secondary eclipse depth underestimates the planetary brightest-region flux, which in turn can lead to underestimating the planetary reflectivity and/or thermal emission. 

Another implication of the phase shift involves the BEER model \citep{faigler11}, developed to search for non-transiting systems through the detection of photometric orbital phase modulations. It assumes that all processes, including gravitational (beaming and ellipsoidal) and atmospheric (reflected light  and thermal emission), are all well aligned with the phase of transit and occultation and that the orbit is circular. Meaning, the extrema of each effect occurs exactly at either transit, occultation, or quadrature phases. When the reflection component is shifted, with either a post- or pre-occultation maximum, then the BEER model will give a biased estimate of the beaming amplitude. This is so since the beaming and reflection components are the sine and cosine components of the same frequency modulation (see \citealt{faigler11} for more details). This in turn will make the companion mass estimate based on the beaming amplitude, $m_{2, beam}$, disagree with the estimate based on the ellipsoidal amplitude, $m_{2, ellip}$. A post-occultation maximum, as identified here, will decrease the beaming amplitude measured by BEER, leading to $m_{2, beam}$ \textless\ $m_{2, ellip}$, while a pre-occultation maximum will lead to $m_{2, beam}$ \textgreater\ $m_{2, ellip}$. 

As noted earlier, the mass discrepancy between the beaming-based mass estimate and the ellipsoidal-based mass estimate has already been measured for several systems, including transiting planets and stellar binaries with a transiting white dwarf companion (see Section~1). Another source of disagreement between the beaming-based and ellipsoidal-based mass estimates are the approximations in the interpretation of the ellipsoidal light curve shape \citep{morris85, morris93, pfahl08}, especially in cases where the primary star rotation is not synchronized with the orbit, or its spin axis is not aligned with the orbital angular momentum axis. Other factors that might be further contributing to this discrepancy are poor understating of the host star \citep[e.g.,][]{shporer14} and non-zero orbital eccentricity. Therefore, deciphering the origin of the discrepancy requires a careful case by case study. \cite{faigler15} recently suggested that the discrepancy for Kepler-13Ab, TrES-2b, HAT-P-7b, and Kepler-76b, identified in \ik\ data, can be attributed to a pre-occultation phase curve maximum of the atmospheric component. They use the difference between the two mass estimates to derive the angular shift of the brightest region on the planetary surface from the substellar point and attribute this shift to asymmetric thermal emission of the planetary atmosphere.

Yet another possible implication of the phase shift and the planetary asymmetric surface brightness distribution is an asymmetry, or a distortion, of the occultation light curve ingress and egress. When identified at sufficiently high signal to noise ratio it can be used to map the planet's surface brightness, a method known as eclipse mapping \citep[e.g.,][]{williams06, dewit12}. When the distortion is not directly identified it can still lead to a shift in the measured occultation phase when modeled using a symmetric occultation light curve model \citep[e.g.,][]{shporer14}. However, the \ik\ data of the three transiting planets studied here is not sensitive to this minute effect. 

Finally, we note that a longitudinally non-uniform atmosphere will affect the shape of the polarization reflected light  phase curve \citep[e.g.,][]{Seager00, Berdyugina11, Madhu12}.

\subsection{Can the atmospheric-induced modulations be used to detect similar but non-transiting exoplanets?}

The three exoplanets studied here (\kepsb, \keptb, and \kepfb) all have a mass of roughly half a Jupiter mass and orbital semi-major axes within 0.1 AU. Such a planet mass gives a gravitational signal (due to the beaming and ellipsoidal effects) at the $\sim$1 ppm level, which is not expected to be detectable in \ik\ data. On the other hand, their \ik\ light curves show a clear atmospheric photometric orbital modulation which if detectable can reveal the existence of a faint low-mass companion even when the system does not transit, as discussed in more details below. Therefore, such planets represent a class of non-transiting planets that although are typically beyond the reach of the BEER approach \citep{faigler11}, as the latter is designed to simultaneously detect all orbital modulations, can potentially still be identified through their atmospheric photometric orbital modulation resulting from their relatively large geometric albedo. The latter is similar to the approach discussed in detail by \cite{jenkins03}. 

Since the overall phased light curve shape is similar to a pure sine a simple Fourier-like period analysis could be used to identify it, at least in principle. This will result in a measurement of the orbital period, and the photometric modulation amplitude will give a mutual constraint on the companion's radius, geometric albedo, and brightness temperature, at the observed wavelength. The companion's (minimum) mass and the true nature of the system can then be obtained via radial velocity follow-up measurements.
Interestingly, \cite{charpinet11} and \cite{silvotti14} used a similar approach to claim the discovery of non-transiting small planets orbiting the hot B subdwarf stars Kepler-70\footnote{Also known as KOI-55, KIC 5807616, and KPD 1943+4058.} and Kepler-429\footnote{KIC 10001893.}, respectively, at short orbital periods. 

\begin{figure*}
\begin{center}
\includegraphics[scale=0.5]{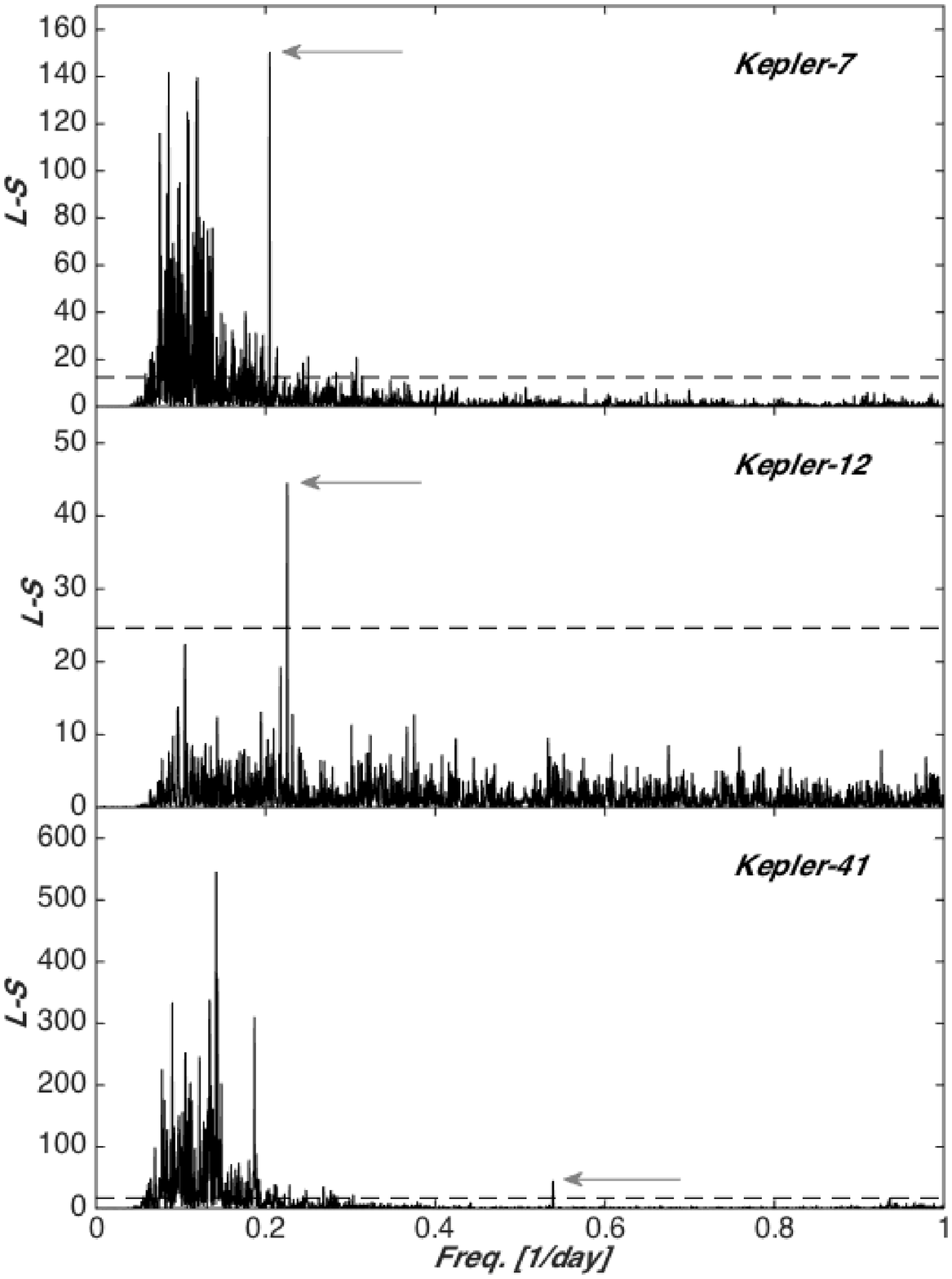}
\caption{\label{fig:per}
Lomb-Scargle (L-S) median-normalized periodograms of \ik\ light curves of \keps\ (top), \kept\ (middle), and \kepf\ (bottom) after removing data during transit and occultation. We used the fast L-S algorithm described by \cite{press89} as implemented in the Matlab/plomb function. In each panel the gray arrow points to the frequency component that coincides with the orbital frequency, and the dashed horizontal black line marks the L-S power level corresponding to a false alarm probability of 10$^{-6}$.
}
\end{center}
\end{figure*}

To see whether \ksb, \ktb, and \kfb\ could have been identified as candidate hot Jupiter systems even if they were not transiting we show in \figr{per} a Lomb-Scargle \citep[L-S;][]{lomb76, scargle82} periodogram of their \ik\ light curves after removing the transit and occultation data. We used the fast L-S algorithm described by \cite{press89} as implemented in the Matlab/plomb function. In all three panels the dashed horizontal black line marks the L-S power level corresponding to a false alarm probability of 10$^{-6}$, and the frequency component at the orbital frequency is marked by a gray arrow.

\kep\ (\figr{per} bottom panel) is known as an active star \citep{quintana13} and its periodogram is dominated by the rotational modulation at frequencies smaller than the orbital frequency. Yet, there is a clear signal at the orbital frequency. For \keps\ and \kept\ (\figr{per} top and middle panels) the strongest frequency components are at the orbital frequency. Meaning, they would have been identified as periodic variables at the orbital period even with no transits. In fact, including the data during conjunctions, which was removed here, would make the orbital frequency component even stronger, as we have confirmed through simulations.

However, there are other processes that may result in photometric sinusoidal variability, so the simplistic approach described above may suffer from a high false positive rate \citep{jenkins03}. One such scenario involves an active star, where the combination of the stellar rotation with the non-uniform spot surface distribution results in sine-like photometric modulations \citep[e.g.,][]{hartman11, mcquillan14}. Although, such modulations vary with time in amplitude and phase, as opposed to a steady orbital signal, so a more sophisticated approach, e.g., using wavelets \citep{bravo14}, can differ between activity-induced modulations and orbital modulations. Moreover, as shown by \cite{mcquillan13} hot Jupiters orbiting at short periods are typically hosted by stars rotating at significantly longer rotation periods, as is the case for \kepfb. So the orbital and rotational frequency components are expected to be well separated in the periodogram. 

Stellar asteroseismic pulsations might also be a false positive scenario, although, typical pulsations of main sequence stars are at a much shorter time scale than hot Jupiters orbital periods \citep[e.g.,][]{chaplin13}. 

Yet another false positive scenario can originate from blending of nearby stars on \ik\ pixels. In that scenario the light from a faint variable star, for example an ellipsoidal variable (contact stellar binary), can be blended with that of a brighter non-variable star to give an observed low-amplitude sinusoidal modulation. This scenario, which presents a problem for {\it any} variability study using \ik\ data, can be examined using light curves of the individual \ik\ pixels \citep{bryson13} combined with high angular resolution imaging of the target that maps the stars within the \ik\ pixel mask \citep[e.g.,][]{horch14, law14}.

Although the detection of hot Jupiters through their atmospheric modulations alone is expected to be more prone to false positive scenarios than BEER, it can allow access to (non-transiting) exoplanets in a lower mass range, down to below a Jupiter mass, where exoplanets are far more common than at a few Jupiter mass or above where they become detectable by the BEER approach \citep{shporer11, faigler13}.

\section{Summary}
\label{sec:sum}

We have identified three \ik\ transiting planetary systems whose light curves are dominated by atmospheric processes, thus allowing a direct view of their atmospheres, independent of any approximations done in the simultaneous modeling of additional orbital effects including beaming and ellipsoidal. We presented the analysis of two of them here, \ktb\ and \kfb, while the third, \kepsb, was analyzed by others \citep{demory13, hu15, webber15}.

Our analysis of \ktb\ and \kepb\ phase curves confirms the light curves' asymmetric nature \citep{esteves15} where the maximum brightness appears post-occultation. Our atmospheric modeling shows that reflective clouds located on the west side of the substellar point can best explain these phase shifts. We hypothesize that the asymmetric cloud coverage is the result of a longitudinal temperature variation driven by zonal jets as the dominant form of atmospheric circulation on those planets. 

The fact that all three systems with atmospheric-dominated \ik\ phase curves show a post-occultation shift of the phase curve maximum suggests it is a common phenomenon, and similar shifts are likely to exist in phase curves of other hot Jupiters where they are more difficult to identify uniquely as they contain additional phase modulations components, induced by beaming and tidal ellipsoidal distortion. 

The phase curve analysis using the atmosphere model framework of \cite{hu15} applied here to \keptb\ and \kepfb\ shows that key properties of clouds in the atmospheres of highly-irradiated exoplanets can be derived from their visible-wavelength phase curves. Further understanding can be obtained by the simultaneous study of phase curves in other wavelength regimes, for example by using \is\ to obtain IR phase curves, or, a dedicated mission like the proposed FINESSE mission \citep{deroo12} to measure phase curves throughout a wide wavelength range from the optical to the IR. 

The significant atmospheric photometric orbital modulation of \kepsb, \keptb, and \kepfb, along with the westward bright-spot shift have possible implications on the search for non-transiting but otherwise similar planets using the BEER approach in \ik\ data \citep{faigler11}, and in high-quality photometric data of other current and future space missions like K2 \citep{howell14}, TESS \citep{ricker14}, and PLATO \citep{rauer14}.

Atmospheric processes (thermal emission and reflected light) resulting in an asymmetric surface brightness are likely to be at play in phase curves showing also the gravitational processes (beaming and tidal ellipsoidal distortion). Such asymmetry can explain some of the mass discrepancy cases, where there is a discrepancy between the beaming-based mass estimate and the ellipsoidal-based mass estimate (see Section~4.2). Therefore it is a step forward in the detailed understanding of visible-light phase curves, and using them through the BEER approach for looking for non-transiting faint low-mass binary companions in short orbital periods \citep{faigler11, faigler15}. However, asymmetric surface brightness cannot explain all the mass discrepancy cases (e.g., KOI-74, \citealt{vankerkwijk10, bloemen12}; KIC 10657664, \citealt{carter11}; see also \citealt{rappaport15}). Hence, fully understanding visible-light phase curves requires further study, especially in cases of hot early-type host stars (like KOI-74 and KIC 10657664) where the ellipsoidal distortion is expected to behave differently than for cooler late-type stars \citep{pfahl08}.

Finally, we have noted that although non-transiting but otherwise similar planets to those analyzed here are not expected to show a detectable gravitational signal, they might still be detectable through their atmospheric photometric orbital modulations. 

\acknowledgments

We are grateful to the anonymous referee whose comments helped improve this paper.
This work was performed in part at the Jet Propulsion Laboratory, under contract with the California Institute of Technology (Caltech) funded by NASA through the Sagan Fellowship Program executed by the NASA Exoplanet Science Institute.
Support for R.H. for this work was provided by NASA through Hubble Fellowship grant \#51332 awarded by the Space Telescope Science Institute, which is operated by the Association of Universities for Research in Astronomy, Inc., for NASA, under contract NAS 5-26555.
This research has made use of NASA's Astrophysics Data System Service.
This paper includes data collected by the \ik\  mission. Funding for the \ik\  mission is provided by the NASA Science Mission directorate.

{\it Facilities: \ik} 


\appendix 

\section{Phase curve figures before and after polynomial detrending}
\label{app:polycomp}

\begin{figure*}
\begin{center}
\includegraphics[scale=0.55]{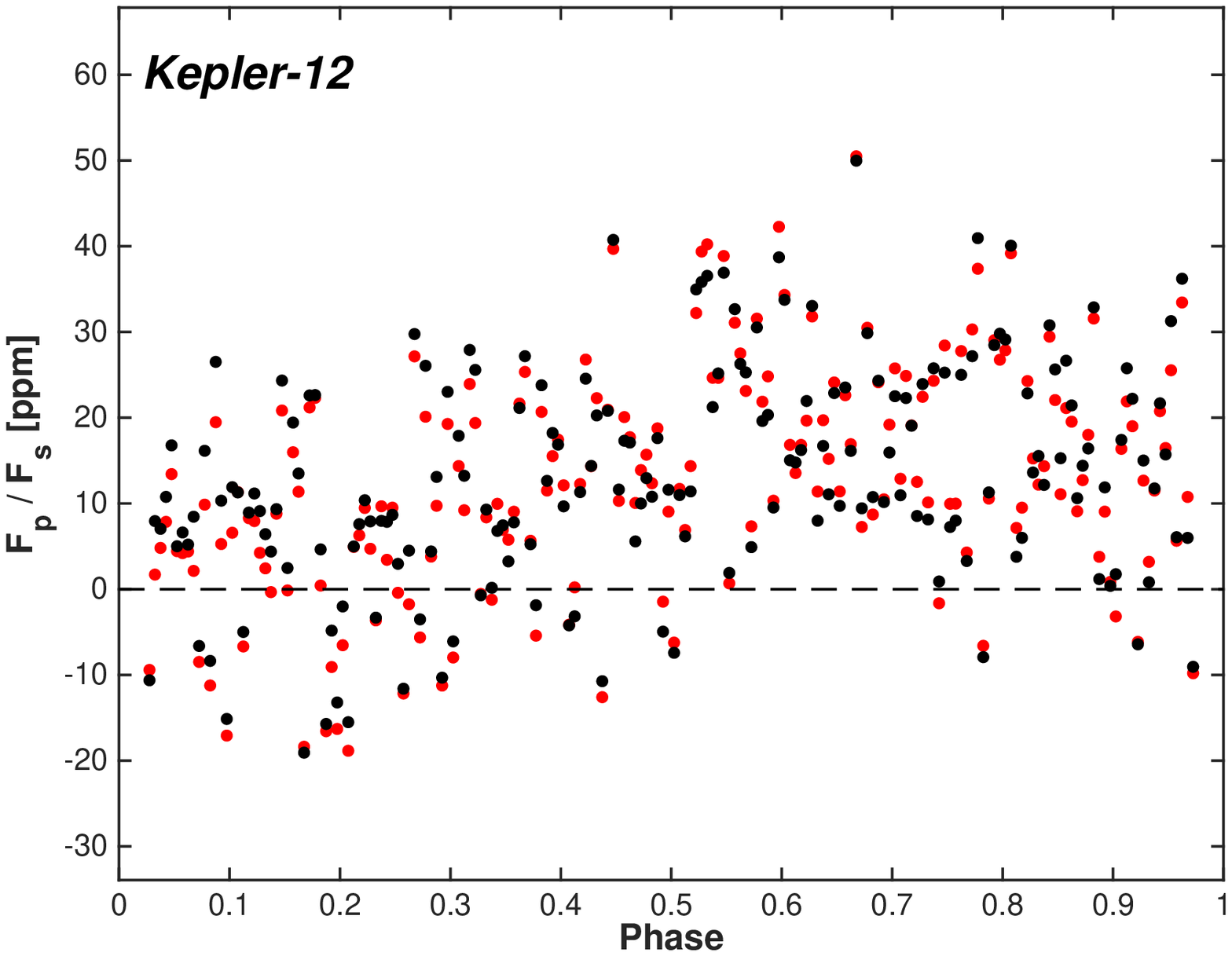}
\includegraphics[scale=0.55]{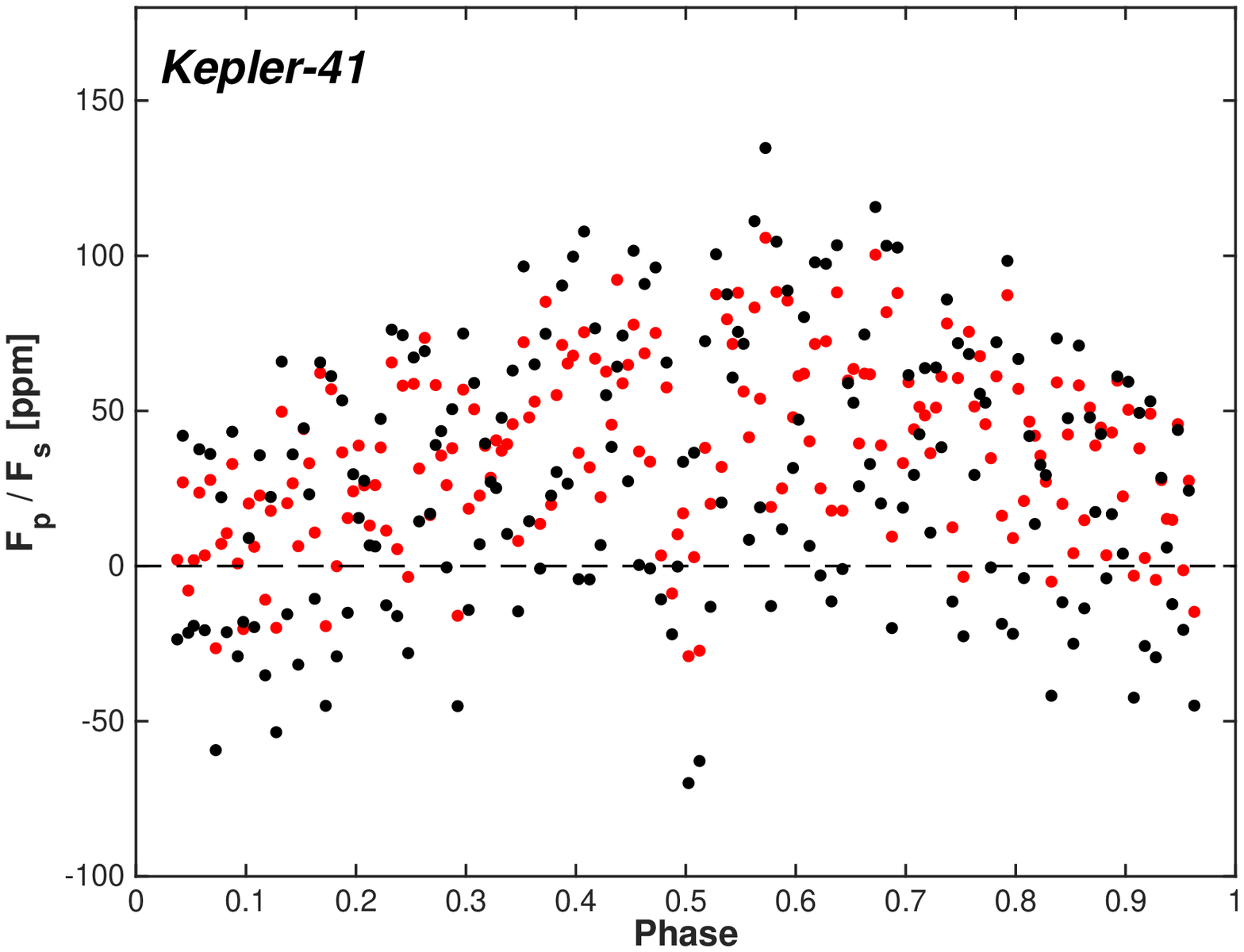}
\caption{\label{fig:polycomp}
Phase curves with (red) and without (black) the polynomial detrending (see \secr{dataanal}), for \kept\ (top) and \kepf\ (bottom). Similarly to \figr{lcmodel} the light curves present the flux from the planet, $F_p$, relative to the in-occultation flux from the host star, $F_s$, in part per million (ppm) versus orbital phase. Error bars are not shown to avoid cluttering the figure.
}
\end{center}
\end{figure*}

\figr{polycomp} presents the two phase curves analyzed here, of \kept\ and \kepf, while showing them with (red) and without (black) the polynomial detrending (see \secr{dataanal}). For \kepf\ it is visually clear that the polynomial detrending decreases the overall scatter, although the difference is minor for \kept. 

\newpage

\section{Correlation matrix figures}
\label{app:matfig}

\begin{figure*}
\begin{center}
\includegraphics[scale=0.6]{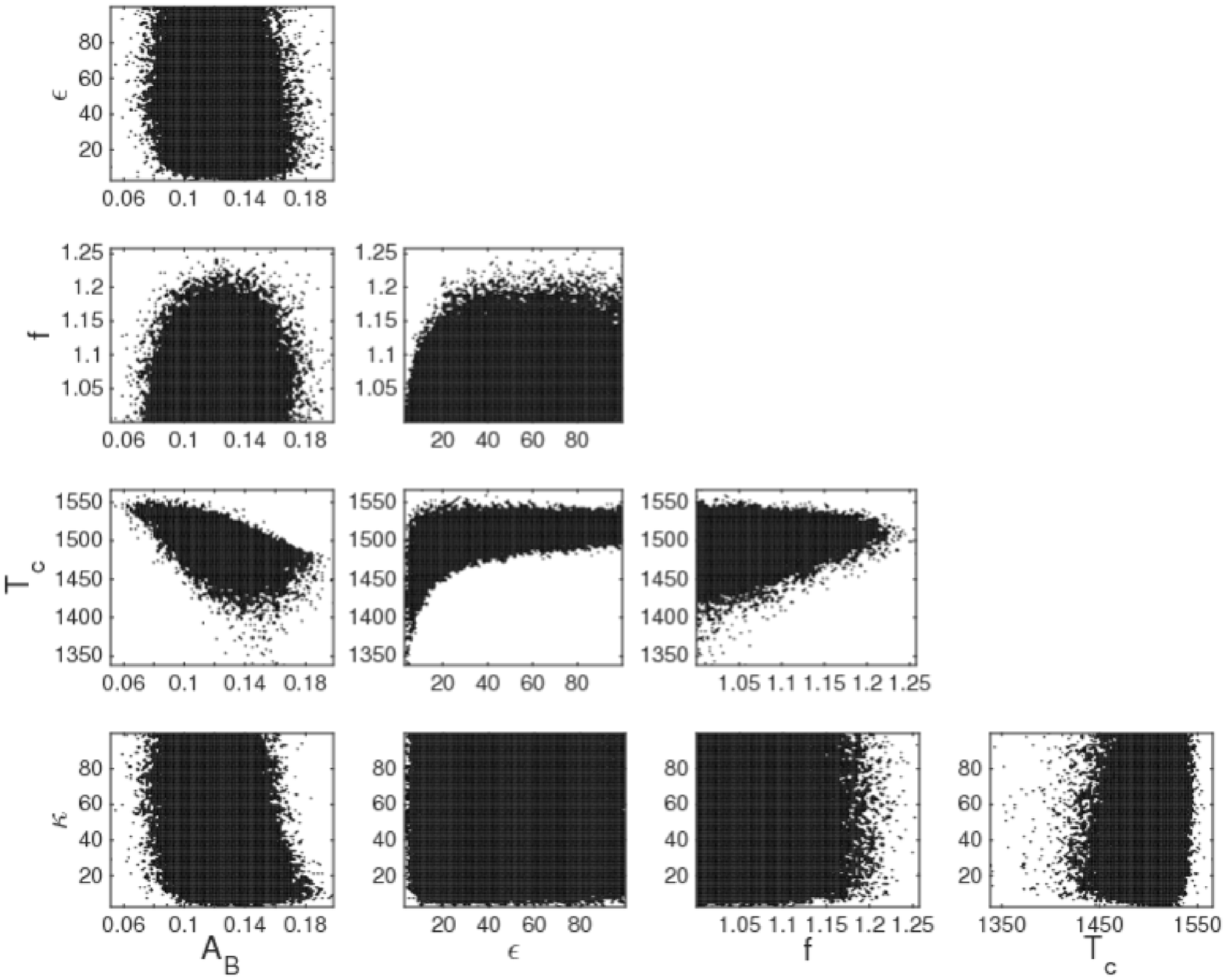}
\caption{\label{fig:corr1}
Correlation matrix for \keptb\ showing the correlation between the five fitted model parameters (see \secr{atm}): Bond albedo ($A_{B}$), heat redistribution efficiency ($\eps$), greenhouse factor ($f$), clouds condensation temperature ($T_{c}$), and clouds reflectivity boosting factor ($\kappa$). 
}
\end{center}
\end{figure*}

\begin{figure*}
\begin{center}
\includegraphics[scale=0.6]{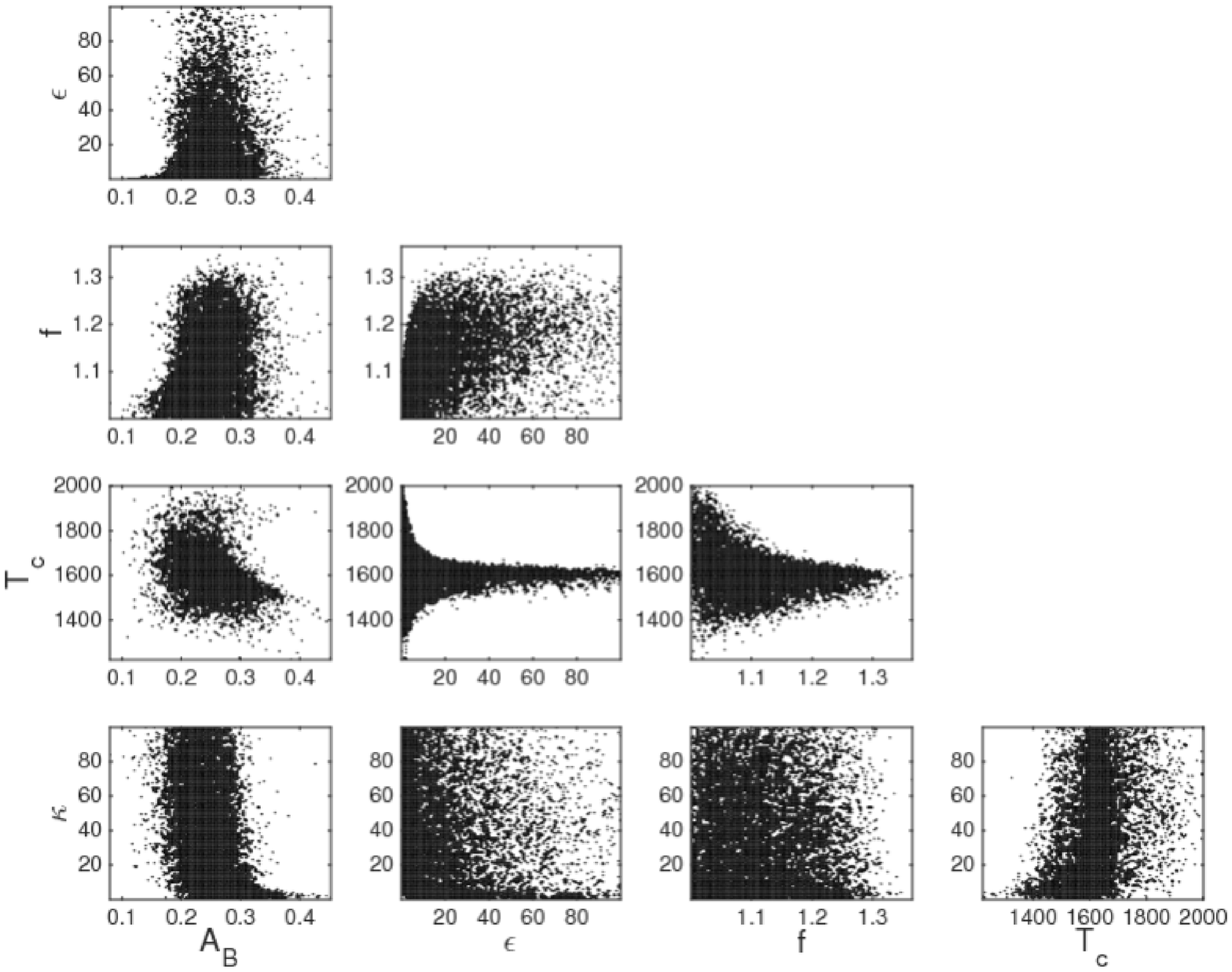}
\caption{\label{fig:corr2}
Same as \figr{corr1}, for \kepfb.
}
\end{center}
\end{figure*}

We present in Figures \ref{fig:corr1} and \ref{fig:corr2} the correlation matrix figures for \ktb\ and \kfb, respectively, where the correlations are shown between any pair of the five fitted model parameters.

\begin{thebibliography}{}

\bibitem[Angerhausen et al.(2014)]{angerhausen14} Angerhausen, D., 
DeLarme, E., \& Morse, J.~A.\ 2014, arXiv:1404.4348 

\bibitem[Barclay et al.(2012)]{barclay12} Barclay, T., Huber, D., 
Rowe, J.~F., et al.\ 2012, \apj, 761, 53 

\bibitem[Berdyugina et al.(2011)]{Berdyugina11} Berdyugina, S.~V., 
Berdyugin, A.~V., Fluri, D.~M., \& Piirola, V.\ 2011, \apjl, 728, L6 

\bibitem[Bloemen et al.(2011)]{bloemen11} Bloemen, S., Marsh, 
T.~R., {\O}stensen, R.~H., et al.\ 2011, \mnras, 410, 1787 

\bibitem[Bloemen et al.(2012)]{bloemen12} Bloemen, S., Marsh, 
T.~R., Degroote, P., et al.\ 2012, \mnras, 422, 2600 

\bibitem[Bouchy et al.(2005)]{bouchy05} Bouchy, F., Pont, F., Melo, C., et al.\ 2005, \aap, 431, 1105 

\bibitem[Bravo et al.(2014)]{bravo14} Bravo, J.~P., Roque, S., Estrela, R., Le{\~a}o, I.~C., \& De Medeiros, J.~R.\ 2014, \aap, 568, A34 

\bibitem[Bryson et al.(2013)]{bryson13} Bryson, S.~T., Jenkins, 
J.~M., Gilliland, R.~L., et al.\ 2013, \pasp, 125, 889 

\bibitem[Carter et al.(2011)]{carter11} Carter, J.~A., 
Rappaport, S., \& Fabrycky, D.\ 2011, \apj, 728, 139 

\bibitem[Chaplin \& Miglio(2013)]{chaplin13} Chaplin, W.~J., \& Miglio, A.\ 2013, \araa, 51, 353 

\bibitem[Charpinet et al.(2011)]{charpinet11} Charpinet, S., 
Fontaine, G., Brassard, P., et al.\ 2011, \nat, 480, 496 

\bibitem[Cowan \& Agol(2011)]{cowan11} Cowan, N.~B., \& Agol, E.\ 2011, \apj, 729, 54 

\bibitem[Cowan et al.(2012)]{cowan12} Cowan, N.~B., Machalek, 
P., Croll, B., et al.\ 2012, \apj, 747, 82 

\bibitem[Demory et al.(2011)]{demory11} Demory, B.-O., Seager, S., Madhusudhan, N., et al.\ 2011, \apjl, 735, L12 

\bibitem[Demory et al.(2013)]{demory13} Demory, B.-O., de Wit, J., Lewis, N., et al.\ 2013, \apjl, 776, L25 

\bibitem[Deroo et al.(2012)]{deroo12} Deroo, P., Swain, M.~R., 
\& Green, R.~O.\ 2012, \procspie, 8442,  

\bibitem[de Wit et al.(2012)]{dewit12} de Wit, J., Gillon, M., Demory, B.-O., \& Seager, S.\ 2012, \aap, 548, A128 

\bibitem[Esteves et al.(2013)]{esteves13} Esteves, L.~J., De 
Mooij, E.~J.~W., \& Jayawardhana, R.\ 2013, \apj, 772, 51 

\bibitem[Esteves et al.(2015)]{esteves15} Esteves, L.~J., De 
Mooij, E.~J.~W., \& Jayawardhana, R.\ 2015, \apj, 804, 150 

\bibitem[Faigler \& Mazeh(2011)]{faigler11} Faigler, S., \& Mazeh, T.\ 2011, \mnras, 415, 3921 

\bibitem[Faigler et al.(2012)]{faigler12} Faigler, S., Mazeh, T., 
Quinn, S.~N., Latham, D.~W., \& Tal-Or, L.\ 2012, \apj, 746, 185 

\bibitem[Faigler et al.(2013)]{faigler13} Faigler, S., Tal-Or, L., Mazeh, T., Latham, D.~W., \& Buchhave, L.~A.\ 2013, \apj, 771, 26 

\bibitem[Faigler \& Mazeh(2015)]{faigler15} Faigler, S., \& Mazeh, T.\ 2015, \apj, 800, 73 

\bibitem[Fortney et al.(2011)]{fortney11} Fortney, J.~J., Demory, B.-O., D{\'e}sert, J.-M., et al.\ 2011, \apjs, 197, 9 

\bibitem[Gelman \& Rubin(1992)]{gelman92} Gelman, A., Rubin, D. B., 1992, Statistical Science, 7, 457

\bibitem[Haario et al.(2006)]{haario06} Haario H., Laine M., Mira A., Saksman E., 2006, Statistics and Computing, 16, 339

\bibitem[Hartman et al.(2011)]{hartman11} Hartman, J.~D., Bakos, 
G.~{\'A}., Noyes, R.~W., et al.\ 2011, \aj, 141, 166 

\bibitem[Horch et al.(2014)]{horch14} Horch, E.~P., Howell, 
S.~B., Everett, M.~E., \& Ciardi, D.~R.\ 2014, \apj, 795, 60 

\bibitem[Howell et al.(2014)]{howell14} Howell, S.~B., Sobeck, 
C., Haas, M., et al.\ 2014, \pasp, 126, 398

\bibitem[Hu \& Seager(2014)]{hu14} Hu, R., Seager, S., 2014, ApJ, 784, 63

\bibitem[Hu et al.(2015)]{hu15} Hu, R., Demory, B.-O., Seager, S., Lewis, N., \& Showman, A.~P.\ 2015, \apj, 802, 51 

\bibitem[Jenkins \& Doyle(2003)]{jenkins03} Jenkins, J.~M., \& Doyle, L.~R.\ 2003, \apj, 595, 429 

\bibitem[van Kerkwijk et al.(2010)]{vankerkwijk10} van Kerkwijk, 
M.~H., Rappaport, S.~A., Breton, R.~P., et al.\ 2010, \apj, 715, 51 

\bibitem[Knutson et al.(2007)]{knutson07} Knutson, H.~A., 
Charbonneau, D., Allen, L.~E., et al.\ 2007, \nat, 447, 183 

\bibitem[Knutson et al.(2009)]{knutson09} Knutson, H.~A., 
Charbonneau, D., Cowan, N.~B., et al.\ 2009, \apj, 703, 769 

\bibitem[Knutson et al.(2012)]{knutson12} Knutson, H.~A., Lewis, 
N., Fortney, J.~J., et al.\ 2012, \apj, 754, 22 

\bibitem[Latham et al.(2010)]{latham10} Latham, D.~W., Borucki, W.~J., Koch, D.~G., et al.\ 2010, \apjl, 713, L140 

\bibitem[Law et al.(2014)]{law14} Law, N.~M., Morton, T., 
Baranec, C., et al.\ 2014, \apj, 791, 35 

\bibitem[Lewis et al.(2013)]{lewis13} Lewis, N.~K., Knutson, 
H.~A., Showman, A.~P., et al.\ 2013, \apj, 766, 95 

\bibitem[Loeb \& Gaudi(2003)]{loeb03} Loeb, A., \& Gaudi, B.~S.\ 2003, \apjl, 588, L117 

\bibitem[Lomb(1976)]{lomb76} Lomb, N.~R.\ 1976, \apss, 39, 447 

\bibitem[Madhusudhan \& Burrows(2012)]{Madhu12} Madhusudhan, N., \& Burrows, A.\ 2012, \apj, 747, 25 

\bibitem[Mazeh \& Faigler(2010)]{mazeh10} Mazeh, T., \& Faigler, S.\ 2010, \aap, 521, L59 

\bibitem[Mazeh et al.(2012)]{mazeh12} Mazeh, T., Nachmani, G., Sokol, G., Faigler, S., \& Zucker, S.\ 2012, \aap, 541, A56 

\bibitem[McQuillan et al.(2013)]{mcquillan13} McQuillan, A., Mazeh, 
T., \& Aigrain, S.\ 2013, \apjl, 775, L11 

\bibitem[McQuillan et al.(2014)]{mcquillan14} McQuillan, A., Mazeh, 
T., \& Aigrain, S.\ 2014, \apjs, 211, 24 

\bibitem[Morris(1985)]{morris85} Morris, S.~L.\ 1985, \apj, 295, 143 

\bibitem[Morris \& Naftilan(1993)]{morris93} Morris, S.~L., \& Naftilan, S.~A.\ 1993, \apj, 419, 344 

\bibitem[Pfahl et al.(2008)]{pfahl08} Pfahl, E., Arras, P., 
\& Paxton, B.\ 2008, \apj, 679, 783 

\bibitem[Press \& Rybicki(1989)]{press89} Press, W.~H., \& Rybicki, G.~B.\ 1989, \apj, 338, 277 

\bibitem[Quintana et al.(2013)]{quintana13} Quintana, E.~V., Rowe, 
J.~F., Barclay, T., et al.\ 2013, \apj, 767, 137 

\bibitem[Rappaport et al.(2015)]{rappaport15} Rappaport, S., 
Nelson, L., Levine, A., et al.\ 2015, \apj, 803, 82 

\bibitem[Rauer et al.(2014)]{rauer14} Rauer, H., Catala, C., Aerts, C., et al.\ 2014, Experimental Astronomy, 38, 249 

\bibitem[Ricker et al.(2014)]{ricker14} Ricker, G.~R., Winn, J.~N., Vanderspek, R., et al.\ 2014, \procspie, 9143, 914320 

\bibitem[Santerne et al.(2011)]{santerne11} Santerne, A., Bonomo, A.~S., H{\'e}brard, G., et al.\ 2011, \aap, 536, A70 

\bibitem[Scargle(1982)]{scargle82} Scargle, J.~D.\ 1982, \apj, 
263, 835 

\bibitem[Seager et al.(2000)]{Seager00} Seager, S., Whitney, 
B.~A., \& Sasselov, D.~D.\ 2000, \apj, 540, 504 

\bibitem[Showman \& Guillot(2002)]{showman02} Showman, A.~P., \& Guillot, T.\ 2002, \aap, 385, 166 

\bibitem[Showman \& Kaspi(2013)]{showman13} Showman, A.~P., \& Kaspi, Y.\ 2013, \apj, 776, 85 

\bibitem[Shporer et al.(2010)]{shporer10} Shporer, A., Kaplan, D.~L., Steinfadt, J.~D.~R., Bildsten, L., Howell, S.~B., \& Mazeh, T.\ 2010, \apjl, 725, L200 

\bibitem[Shporer et al.(2011)]{shporer11} Shporer, A., Jenkins, J.~M., Rowe, J.~F., et al.\ 2011, \aj, 142, 195 

\bibitem[Shporer et al.(2014)]{shporer14} Shporer, A., O'Rourke, 
J.~G., Knutson, H.~A., et al.\ 2014, \apj, 788, 92 

\bibitem[Silvotti et al.(2014)]{silvotti14} Silvotti, R., Charpinet, S., Green, E., et al.\ 2014, \aap, 570, AA130 

\bibitem[Southworth(2008)]{southworth08} Southworth, J.\ 2008, 
\mnras, 386, 1644 

\bibitem[Steinfadt et al.(2010)]{steinfadt10} Steinfadt, J.~D.~R., 
Kaplan, D.~L., Shporer, A., Bildsten, L., \& Howell, S.~B.\ 2010, \apjl, 716, L146 

\bibitem[Still \& Barclay(2012)]{still12} Still, M., \& Barclay, T.\ 2012, Astrophysics Source Code Library, 8004 

\bibitem[Sudarsky et al.(2000)]{sudarsky00} Sudarsky, D., Burrows, A., Pinto, P. 2000, ApJ, 538, 885

\bibitem[Sudarsky et al.(2003)]{sudarsky03} Sudarsky, D., Burrows, A., Hubeny, I., 2003, ApJ, 588, 1121

\bibitem[Webber et al.(2015)]{webber15} Webber, M.~W., Lewis, N.~K., Marley, M., et al.\ 2015, arXiv:1503.01028 

\bibitem[Williams et al.(2006)]{williams06} Williams, P.~K.~G., 
Charbonneau, D., Cooper, C.~S., Showman, A.~P., \& Fortney, J.~J.\ 2006, \apj, 649, 1020 

\bibitem[Winn et al.(2008)]{winn08} Winn, J.~N., Holman, 
M.~J., Torres, G., et al.\ 2008, \apj, 683, 1076 

\bibitem[Zucker et al.(2007)]{zucker07} Zucker, S., Mazeh, T., \& Alexander, T.\ 2007, \apj, 670, 1326 

\end{thebibliography}
\end{document}